\documentclass{appolb}
\usepackage{graphicx}
\usepackage{cite}
\usepackage{soul}   
\begin{document}
\title{The REDTOP experiment: a $\eta/\eta^{\prime}$ factory to explore dark matter and physics beyond the Standard Model%
\thanks{Presented at Workshop at 1 GeV scale: From meson to axions, Krak\'{o}w (Poland), September 19-20, 2024}%
}
\author{Marcin Zieliński
\address{M. Smoluchowski Institute of Physics, Jagiellonian University in Kraków, Poland\\}
%{Corrado Gatto % of different affiliation
%\address{XX}
%}
\vspace{4mm}
Corrado Gatto
\address{Istituto Nazionale di Fisica Nucleare, Italy and Northern Illinois University, USA}
\vspace{4mm}\\
for the REDTOP Collaboration
}

\maketitle

\begin{abstract}
The REDTOP experiment is a proposed super-$\eta$/$\eta'$ factory aimed at exploring physics beyond the Standard Model in the MeV-GeV range and rare $\eta$/$\eta'$ meson decays. With projected production rates exceeding $10^{13}$ $\eta$/year and $10^{12}$ $\eta'$/year, REDTOP will enable studies of symmetry violations and all four portals to the Dark Sector. Preliminary studies show sensitivity, which could open a broad possibility for exploring new physics and contribute to a deeper understanding of fundamental interactions within the Standard Model.
Such high statistics experiments and the required sensitivity can only be achieved with a high intensity proton or pion beam, available at several accelerator facilities worldwide. This article discusses the physics potential of the REDTOP experiment, the detector design, and the future beam requirements.
\end{abstract}
  
\section{Introduction}\label{SectionIntro}
The Standard Model (SM) of particles provides an internally consistent and experimentally validated description of fundamental interactions~\cite{Pais:1975gn}. It is a comprehensive framework for describing the electromagnetic, weak, and strong interactions of elementary particles~\cite{Wilczek:2004rm}. However, a growing number of experimental observations suggest that it may be incomplete~\cite{Reece:2009un}. These include not only long standing puzzles such as the nature of Dark Matter and the origin of neutrino masses, but also more recent measurements on the muon anomalous magnetic moment \((g-2)_{\mu}\)~\cite{Muong-2:2021ojo}, studies of the electric dipole moment (EDMs)~\cite{Purcell:1950zz,Abel:2017rtm}, and rare decay anomalies~\cite{Krasznahorkay:2015iga,Darme:2022zfw}, which may point toward new physics at \textbf{st} intermediate energy scales. Moreover, the absence within the SM of CP violation sources sufficient to explain baryogenesis, together with the lack of a Dark Matter candidate, are a strong hint of the existence of new weakly coupled particles. 
Many  open cases motivate the search for physics Beyond the Standard Model (BSM) through both high energy collisions and precision measurements in rare or forbidden processes. In this context, experiments with high intensity beams searching for particles at 1~GeV scale could be essential to uncover new physics. 
Currently REDTOP ({\bf{R}}are {\bf{E}}ta {\bf{D}}ecays {\bf{TO}} {\bf{P}}robe New Physics) is one of the dedicated experimental initiatives specifically conceived to investigate rare processes in 1~GeV energy regime, where new particles or violations of fundamental symmetries may emerge. A central focus of the REDTOP scientific program is the investigation of rare decays of $\eta$ and $\eta^{\prime}$ mesons produced by fixed targets with proton and pion beams at energies of a few GeV~\cite{REDTOP:2022slw}. By focusing on channels that are highly suppressed within the Standard Model, REDTOP aims to achieve a sensitivity improvement of several orders of magnitude over previous efforts~\cite{WASA-at-COSY:2004mns,Amelino-Camelia:2010cem}. Such an advancement is crucial for accessing potential manifestations of phenomena that remain beyond the reach of current high energy collider experiments, thus establishing REDTOP as a complementary approach in the broader search for extensions of the Standard Model. In this article, we will present the key theoretical concepts underlying the REDTOP experimental program, followed by a detailed description of the detector design, with a particular focus on its most critical components essential for the planned measurements.

\section{Theoretical framework}\label{SectionFramework}
Modern theoretical models points out that potential interactions between New Physics and the Standard Model are expected to occur with extremely small coupling constants, typically on the order of 10\(^{-8}\) or lower~\cite{Batell:2009di,alexander2016darksectors2016workshop,essig2013darksectorsnewlight}. 
These models involving hidden sectors are often characterized by the presence of so called portals~\cite{Batell:2009di}, where a new field, such as a vector boson, scalar, pseudoscalar, or heavy neutral lepton interacts with the Standard Model via a gauge singlet interaction of mass dimension four or lower. These portals provide a way for new physics to interact with the SM without disrupting its UV properties, thus maintaining the renormalizability of the Standard Model. Additionally, portals with higher mass dimensions, such as those in axion models, offer further potential for exploring new phenomena beyond the SM. 
In this context, the investigation of processes involving particles that do not carry Standard Model charges becomes particularly relevant for Light Dark Matter (LDM) searches. 
%\st{This is because the absence of charged currents eliminates the possibility of interference from Standard Model processes, allowing for a clearer and more direct probe of potential BSM interactions.}
The absence of SM charges and, consequently, of charged current interactions,
minimizes background and interference from known processes, enabling a cleaner and more direct probe of potential Beyond the Standard Model (BSM) interactions.

In line with the requirement for LDM to remain electrically neutral, the $\eta$ and $\eta^{\prime}$ mesons stand out as particularly relevant particles in this context. They are members of the pseudoscalar nonet and play an important role in understanding low energy QCD. Both have isospin and angular momentum equal to zero, negative parity, and charge conjugation equal to +1 ($I^G(J^{PC}) = 0^+(0^{-+}$))~\cite{ParticleDataGroup:2024cfk}. Therefore, like the Higgs boson and the vacuum, they carry no Standard Model charges, a unique property that allows them to interact in ways that do not interfere with SM processes. Moreover, all $\eta$ and $\eta^{\prime}$ decays are flavor-conserving, with small decay widths $\Gamma_{\eta} = 1.31$~keV and  $\Gamma_{\eta^{\prime}} = 0.188$~MeV which makes them a perfect candidate for the search for small BSM effects. Also, all electromagnetic and strong decays are suppressed up to order $\mathcal{O} (10^{-8})$ favoring the exploration of more rare processes. 

To explore portals and their role in probing hidden sectors, several $\eta$ and $\eta^{\prime}$ related processes have been identified as key for investigating these new interactions. 

\textit{The Vector portal.}
The Vector portal encompasses a broad range of theoretical models in which a new vector mediates interactions between the Standard Model and hidden sectors.
The most accredited models are: Minimal dark photon model~\cite{Holdom:1985ag},  Leptophobic B boson Model~\cite{Tulin:2014tya,Escribano:2022njt}, and Protophobic Fifth Force model~\cite{Feng:2016jff,Feng:2016ysn}. They can be investigated through radiative decays $\eta\to\gamma A^{\prime}$, which may subsequently decay into a lepton-antilepton pair \(l^+l^-\) or two pions. 
%\st{In contrast, the scalar or Higgs, portal models, interacts with the SM through couplings to the Higgs boson or its extensions. }

\textit{The Scalar portal.}
The scalar portal can be probed through the decay channels of the $\eta$ meson that involve the production of a neutral pion $\pi^{0}$  in association with lepton or pion pairs through the decay channels: 
\(\eta\to\pi^{0}H\to \pi^{0}l^{+}l^{-}\) or \(\eta\to\pi^{0}H\to \pi^{0}\pi^{+}\pi^{-}\). 
Three complementary scalar portal models are presently being explored: the Minimal Dark Scalar Model, the Spontaneous Flavor Violation Model\cite{Egana-Ugrinovic:2018znw}, and the Flavor-Specific Scalar Model, which display similar experimental signatures along with the Two-Higgs Doublet Model~\cite{Batell:2021xsi,Abdallah:2020vgg}. 
In the Minimal Dark Scalar Model, the dark scalar corresponds to a light Higgs boson and predominantly couples to heavy quarks, whereas in the Two-Higgs Doublet Model, the coupling is enhanced for light quarks~\cite{Batell:2018fqo}. This distinction results in decay branching ratios that differ by more than two orders of magnitude, allowing to differentiate between the two models.

\textit{The Pseudoscalar portal.}
The pseudoscalar portal constitutes a particularly rich sector for exploring BSM physics, with several recent theoretical models predicting the existence of new light pseudoscalar states. In particular, the long sought but still unconfirmed Peccei–Quinn mechanism introduces the axion~\cite{Peccei:1977hh,Peccei:1977ur}, a hypothetical particle originally proposed to resolve the strong CP problem in QCD. In addition to its theoretical appeal, Axion Like Particles (ALPs)~\cite{Bauer:2021wjo} have recently gained renewed interest due to their potential to explain anomalies observed in various experimental results\cite{Krasznahorkay:2015iga}, further motivating dedicated searches in rare $\eta$ and $\eta^{\prime}$ meson decays~\cite{Alves:2017avw}. Among the possible signatures, axion-hadronic decays resulting in two charged or neutral pions are of special interest. In this scenario, GeV scale dynamics coupling to the first generation of Standard Model fermions generates a short lived QCD axion or an ALP that decays predominantly into $e^+e^-$ pairs. In this context, decays such as \(\eta^{(\prime)} \to \pi\pi a \to \pi\pi e^{+}e^{-}\) would constitute a critical test for the existence of a pseudoscalar state coupled to quarks and gluons~\cite{Alves:2020xhf,Alves:2024dpa}. Such a coupling would alter the QCD topological vacuum and affect the strong CP phase \(\theta_{{QCD}}\). Although an ALP could contribute to \(\theta_{{QCD}}\) without resolving the CP problem, which requires additional fine-tuning, the detection of a pseudoscalar capable of dynamically \(\theta_{{QCD}}\) forcing to zero would provide compelling evidence for the QCD axion itself.

\textit{The  Heavy Neutral Lepton portal.}
The heavy neutral lepton portal (HNL) introduces one or more dark fermions that mix with Standard Model neutrinos, offering a natural framework for addressing neutrino masses and BSM phenomena. In the current models, attention is primarily given to realizations within the Two-Higgs Doublet Model (2HDM), which predicts distinctive signatures in rare $\eta$ and $\eta^{\prime}$ meson decays~\cite{Abdallah:2020vgg}. The specific process explored involves decays $\eta/\eta^{\prime} \to \pi^{0} H$, followed by $H \to \nu N_2$ and subsequent transitions $N_2 \to h^{\prime} N_1$ with $h^{\prime} \to e^+e^-$. Within the Two-Higgs Doublet Model framework and under the assumption of $\lambda_u = \lambda_d$, the predicted branching ratio for these channels is on the order of $\mathcal{O}(10^{-13})$. But when $\lambda_u \neq \lambda_d$ the branching ratios for $H$ along with those for $N_2$ and $h^{\prime}$, are at the level of 10\(^{-12}\), they are within the reach of the REDTOP experiment. While the heavy neutral lepton portal offers an intriguing opportunity for probing new physics, the sensitivity levels required are extremely challenging and currently lie beyond the reach of present day experiments. 

In addition to searches for new particles via hidden portals, precision studies of fundamental symmetries represent a complementary and equally important approach to probing physics beyond the Standard Model. 
Investigations of discrete symmetries, such as $C$, $P$, and $CP$, through rare $\eta$ and $\eta^{\prime}$ decays offer unique sensitivity to potential symmetry violations and provide an essential test bench for many BSM scenarios. Several $\eta$ and $\eta^{\prime}$ decay channels have been identified as crucial for testing fundamental conservation laws. Current studies focus primarily on CP violation, lepton flavor universality, and lepton flavor violation, as these phenomena are supported by well motivated theoretical models. One of the well known approach to investigate C and CP violation with $\eta$ meson  is the study of mirror asymmetries in the Dalitz Plot of the $\eta \to \pi^+\pi^-\pi^0$ decay~\cite{Gardner:2019nid,Akdag:2021efj}. In this process, the interference between a C-conserving but isospin breaking amplitude and a C-violating amplitude would generate a charge asymmetry in the Dalitz plot of the $3\pi$ final state. Given that parity ($P$) is conserved in $\eta \to \pi^+\pi^-\pi^0$ decays, the observation of a non zero charge asymmetry would constitute clear evidence for the violation of both C and CP symmetries. Also, C and CP can be studied using the Dalitz Plot asymmetries in the $t$ and $u$ Mandelstam variables~\cite{Layter:1972aq,Zielinski:2012lsn}. In addition, CP violation can also be investigated through polarization studies of the virtual photon in $\eta \to \pi^+\pi^-\gamma^* \to \pi^+\pi^- e^+e^-$~\cite{Gao:2002gq,Herczeg:1973ese}.
In this decay, the internal conversion of the photon into an $e^+e^-$ pair allows the CP-violating effects, encoded in the polarization of the virtual photon, to manifest as an asymmetry in the angular correlation between the di-lepton ($e^+e^-$) and di-pion ($\pi^+\pi^-$) planes. The observation of a non vanishing transverse polarization component would thus constitute clear evidence of CP-violating phenomena.

All processes mentioned above not only provide critical insight into the physics of hidden sectors but may also help clarify anomalies observed in recent experiments, offering a unique opportunity to test the boundaries of the Standard Model.

\section{The experimental concept and detector requirements}\label{SectionDetector}
Given the nature of LDM and BSM, experiments utilizing fixed-target setups with low-energy, high intensity beams emerge as crucial tools in probing these subtle phenomena. Such experiments provide the necessary sensitivity to detect signals that would otherwise remain hidden in high energy collider experiments, where the focus is on much more energetic processes. 
Nowadays, to effectively explore this broad spectrum of phenomena, $\eta$ and $\eta^{\prime}$-factories are essential for collecting large statistics of events in the MeV-GeV energy range. 
Given the current constraints on BSM parameters and the limitations of available detector technologies, REDTOP aims to produce at least 10\(^{14}\) $\eta$ mesons and 10\(^{12}\) $\eta^{\prime}$ mesons. Achieving such statistics would significantly enhance the physics reach of the experiment, enabling detailed investigations across multiple aspects of BSM physics.

\begin{figure}[!h]
  \centering
  \includegraphics[width=1.0\textwidth]{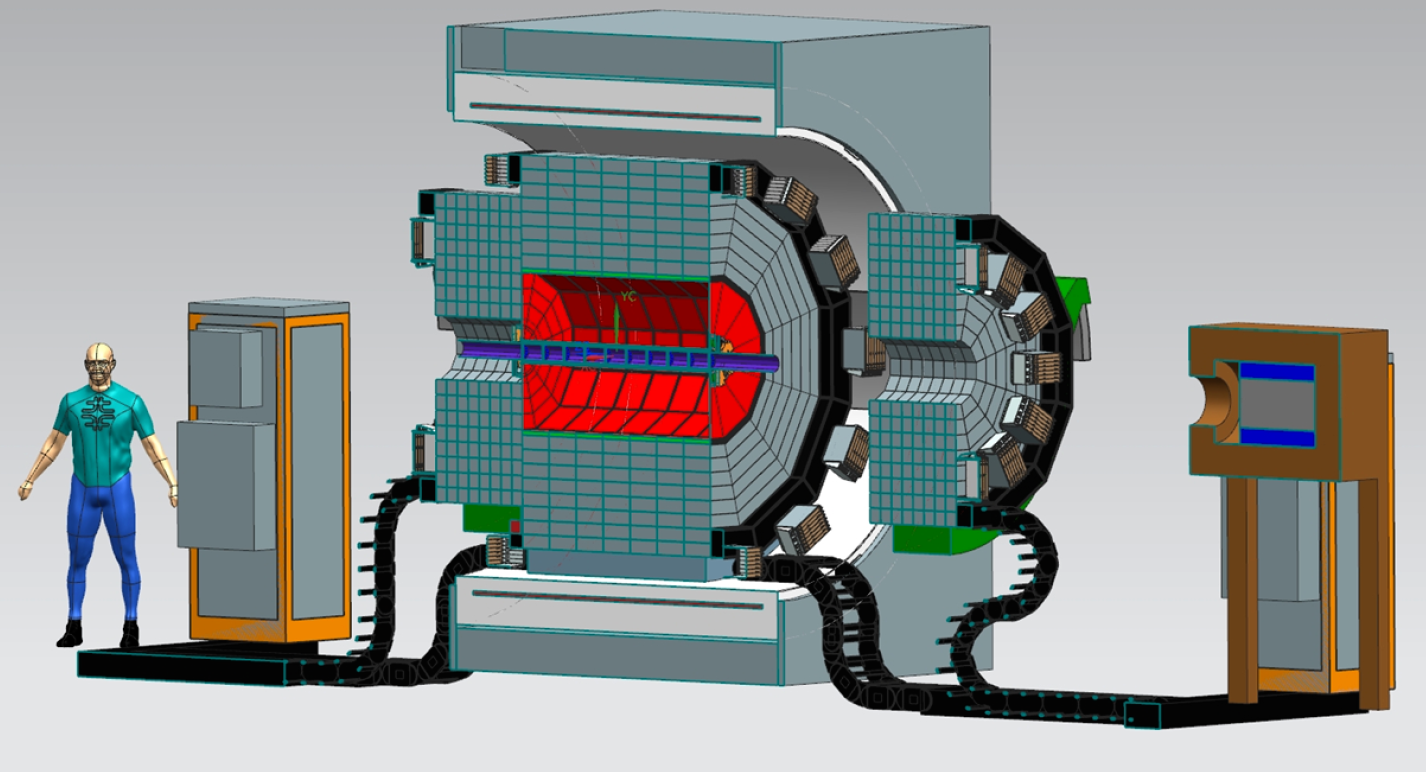}
  \caption{ Schematic layout of the REDTOP detector. }
  \label{fig:reddetector}
\end{figure}

The REDTOP detector has been designed to meet the specific experimental challenges associated with the detection of rare $\eta$ and $\eta'$ decays. In particular, the sensitivity to weakly coupled hidden sector states, CP-violating observables, and lepton universality violating processes demands a detector with excellent vertex resolution, a low material budget, high lepton identification efficiency, and minimal hadronic background. 
To address these challenges, REDTOP adopts a novel detector concept optimized for the efficient tagged production and reconstruction of $\eta$ and $\eta'$ decays in a high intensity proton or pion beam environment. The conceptual layout of the REDTOP detector is shown in Fig.~\ref{fig:reddetector}. 
One of the challenges is to cope with the high inelastic interaction rate, approaching 1~GHz, which will require fast timing capabilities and high spatial granularity to reduce event pileup. 
Moreover, the interaction process is dominated by baryon production, primarily slow moving protons and neutrons, which will contribute significantly to the background. Therefore, REDTOP is equipped with an efficient tracking system capable of disentangling particles with low momentum. 
Finally, most of the channels of interest involve leptons and photons in the final state, making it essential to implement a detector with excellent particle identification. 
In particular, a double-layer Cherenkov Threshold Detector (CTOF) is employed to assist in lepton identification and to provide further rejection of hadronic background via time-of-flight and threshold detection of fast particles. 
An all-silicon tracking system, the ADRIANO3 triple-readout calorimeter, and an optional active muon polarimeter are also important components of the proposed apparatus. 
The REDTOP detector will operate within a large superconducting solenoid, providing a uniform solenoidal magnetic field of 0.6~T that is essential for precise momentum reconstruction and charged particle tracking. 
Together, these components contribute to a detector system with near-$4\pi$ acceptance, a key requirement for the comprehensive detection of all final state particles.

The performance of the REDTOP detector reflects the ambitious physics goals and the challenging experimental conditions associated with $\eta/\eta^{\prime}$ hadro-production. The general detector requirements include calorimeter energy resolution at the level of $\sigma(E)/E \approx 2{--}3\%/\sqrt{E}$, particle identification (PID) efficiencies exceeding 98--99\% for electrons and photons, and around 95\% for muons and charged pions. The required efficiencies for the identification of protons and neutrons reach as high as 99.5\%. The timing resolutions are expected to be 30~ps for the tracking system, 80~ps for the calorimeter, and 50~ps for time-of-flight measurements. These specifications are essential to ensure the precise reconstruction of rare decay topologies, suppression of backgrounds from  baryon production, and accurate timing and PID in a high interaction rate environment. 
To achieve such parameters, each detection subsystem requires a novel technological approach that is currently undergoing active R\&D work. 

\section{Beam requirements}\label{SectionBeam}
The realization of the REDTOP physics program relies on the availability of a high intensity proton beam (or, as an alternative, a pion beam)  with kinetic energies in the range of 1.8–3.5~GeV and an intensity of at least $10^{11}$ protons on target per second (POT), corresponding to an integrated annual exposure of approximately $10^{18}$~POT. To reach this yield of $10^{13}$ $\eta$ mesons annually, equivalent to $10^6$ $\eta$/s, an inelastic interaction rate of about $10^8$ proton-target collisions per second is required. This rate can be achieved using a thin target composed of lithium or beryllium, with a thickness corresponding to $2 \times 10^{-2}$ interaction lengths (approximately 7.7~mm or 2.3~mm, respectively). 

Under these conditions, the total event rate observed by the 
detector is estimated to be on the order of $7\times 10^8$~Hz. Several accelerator facilities worldwide possess the technical capability to deliver such a beam. Assuming successful implementation, this configuration would result in an annual yield of approximately $10^{13}$ $\eta$ and $10^{12}$ $\eta^\prime$ mesons. 

A pion beam  with kinetic energies in the range of 0.8–1.5~GeV and the same intensity indicated above would provide a similar sample of $\eta/\eta^{\prime}$ mesons, but with only 1/4 of the QCD background. 
The events from such a background have a topology that is much easier to recognize and reject than when using a proton beam, with an overall gain in signal-to-noise ratio (S/B) of almost one order of magnitude.
In particular, when a $\pi^-$ beam is used on a Li/Be ($He_3$) target, the $\eta/\eta^{\prime}$ are produced in association with   a neutron (tritium).
If a tagging detector is used, the $\eta/\eta^{\prime}$ are fully tagged, portending an even higher S/B ratio.
In conclusion, by operating the REDTOP experiment with a pion beam, the increase in sensitivity to LDM and BSM physics would be increased by a factor $\approx\sqrt{10}$.

Pion beams are more complex and expensive to implement, requiring an intermediate production target and a pion collector system.
The latter have large inefficiencies, requiring a primary proton beam with an energy above 1.3 GeV and an intensity of the order of $\approx$ 100 KW.
The only two laboratories presently capable of providing such a beam are the ESS (Sweden) and Oak Ridge Natl. Lab (USA).

\section{Conclusions and outlook}\label{SectionSummary}
The REDTOP experiment provides a promising opportunity to investigate New Physics, rare decays, and processes that violate symmetries in the MeV-GeV mass range. By combining high meson yields, near-$4\pi$ acceptance, and new detector technologies, REDTOP can achieve precision measurements with sensitivities that were previously unattainable. The REDTOP proposal addresses several important physics cases, such as QCD axions and axion-like particles, heavy neutral leptons, and tests of fundamental symmetries like C, CP, and lepton flavor violation. The proposed production rates of $\eta$ and $\eta'$ mesons for REDTOP maximize its sensitivity to a wide range of BSM and LDM phenomena. REDTOP seeks to significantly improve the sensitivity and precision of measuring different decay rates by several orders of magnitude (branching ratios up to \(10^{-12}\)) compared to previous experiments.

\section*{Acknowledgments}
We acknowledge support by the National Science Centre, Poland (NCN) through grant No. 2023/50/E/ST2/00673, and the Jagiellonian University's own funds. 
%%%%
\bibliographystyle{unsrt} % styl bibliografii
\bibliography{apbib} % bez .bib

\end{document}